\documentclass[epj]{svjour}
\usepackage{graphics,graphicx,dcolumn,bm,fleqn,epic,eepic,float}
\usepackage{amssymb,amsmath,multirow,rotate,color,float,times}
\usepackage{color}
\usepackage{soul}                    
\definecolor{red}{rgb}{1,0,0}
\definecolor{green}{rgb}{0,1,0}
\definecolor{blue}{rgb}{0,0,1}

\newcommand{\bite}{\begin{itemize}}
\newcommand{\eite}{\end{itemize}}
\newcommand{\benu}{\begin{enumerate}}
\newcommand{\eenu}{\end{enumerate}}
\newcommand{\bverb}{\begin{verbatim}}
\newcommand{\everb}{\end{verbatim}}

\newcommand{\beq}{\begin{equation}}
\newcommand{\eeq}{\end{equation}}
\newcommand{\barr}{\begin{array}}
\newcommand{\earr}{\end{array}}

\begin{document}
\title{The dynamics of financial stability in complex networks}
\author{Jo\~ao P.~da Cruz\inst{1,2} \and Pedro G.~Lind\inst{1,3}
}                     
\offprints{}          
\institute{Departamento de F\'{\i}sica, Faculdade de Ci\^encias 
             da Universidade de Lisboa, 1649-003 Lisboa, Portugal 
          \and 
          Closer Consultoria Lda, Avenida Engenheiro Duarte Pacheco,
             Torre 2, 14$^o$-C, 1070-102 Lisboa, Portugal
           \and
          Center for Theoretical and Computational Physics, 
             University of Lisbon,
             Av.~Prof.~Gama Pinto 2, 1649-003 Lisbon, Portugal}
\date{Received: date / Revised version: date}
%
\abstract{%
We address the problem of banking system resilience by applying off-equilibrium 
statistical physics to a system of particles, representing the economic agents, 
modelled according to the theoretical foundation of the current banking regulation, the so called 
Merton-Vasicek model. Economic agents are attracted to each other to exchange `economic energy', 
forming a network of trades. When the capital level of one economic agent drops below a minimum, 
the economic agent becomes insolvent. The insolvency of one single economic agent affects the 
economic energy of all its neighbours which thus 
become susceptible to insolvency, being able to
trigger a chain of insolvencies (avalanche). 
We show that the distribution of avalanche sizes 
follows a power-law whose exponent depends on the minimum capital level. Furthermore, we present evidence 
that under an increase in the minimum capital level, large crashes will be avoided only if one assumes 
that agents will accept a drop in business levels, while keeping their trading attitudes and 
policies unchanged. The alternative assumption, that agents will try to restore their business 
levels, may lead to the unexpected consequence that large crises occur with higher probability.} 
\maketitle

\section{Introduction}
\label{sec:introduction}
The well-being of humankind depends crucially on the financial stability of the underlying economy. 
The concept of financial stability is associated with the set of 
conditions under which the process of financial 
intermediation (using savings from some economic agents to lend to other economic agents) is smooth, thereby
promoting the flow of money from where it is available to where it is needed. This flow of money is made 
through economic agents, commonly called `banks', that provide the service of intermediation and an upstream flow 
of interest to pay for the savings allocation. 
Because the flow of money that ensures financial stability occurs on top of a complex interconnected set of economic agents 
(network), it must depend not only in individual features or conditions imposed to the economic agents but also on the overall 
structure of the entire economic environment. The role of
banking regulators is to protect the flow of money through the system by implementing rules 
that insulate it against individual or localised breaches that happen when a bank fails to pay back to depositors. 
However, these rules do not always take into account the importance of the topological structure of the network
for the global financial stability. In this paper, we will present quantitative evidence that neglecting the 
topological network structure when implementing financial regulation may have a strong negative impact on financial stability.
\begin{figure}[t]
\centerline{\includegraphics[width=0.90\linewidth]{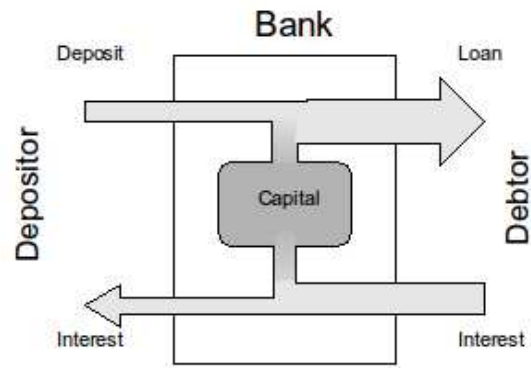}}
\vspace*{8pt}
\caption{\protect
         Illustration of a bank `apparatus' for money flow. A bank lends money to 
	  debtors using money from depositors and also its own money, the capital. 
	  In return debtors pay interest to the bank, which keeps a part to itself
	  and pays the depositors back.}
\label{fig:capital}
\end{figure}

The event of not paying back the money owed is called `default'. In order that downstream defaults do not generate the 
default of a particular  bank, each bank holds an amount of money as a reserve for paying back its depositors. 
In other words, a part of 
the money one bank sends downstream is its own money. 
This share of own money is called `capital' (see Fig \ref{fig:capital}).
Looking to one single bank, if it has a large amount of capital, one 
reasonably expects that the bank will 
also cover a proportionally large debtor default, guaranteeing the deposits made 
by its depositors. On the contrary,
if the capital level of the bank is small, a small debtor default 
is sufficient to put the bank with no conditions for guaranteeing the money
of all its depositors. Loosing such conditions, the bank enters a situation 
called bankruptcy or insolvency.
Usually, bank regulators base their rules in such arguments.

In 1988, a group of central bank governors called the Basel Committee on Banking Regulation 
unified the capital level rules that were applied in each of the member countries and defined 
a global rule to protect the banking system that was becoming global at the time\cite{basel1}. 
Roughly speaking, these rules imposed a minimum capital level of 8\% without any empirical 
reason. A few years later the accord came under criticism from market agents who felt that it did not 
differentiate enough between the various debtors, i.e.~between the entities
whom the bank lends money, and a second version of the accord\cite{basel2} was 
finished in 2004 to become effective in 2008. In this second version, banks were allowed to 
use the Merton-Vasicek model\cite{MertonVasicek} based on the Value-at-Risk paradigm(VaR)\cite{varref} to 
weight the amount of lent money in the calculation of the 
necessary capital according to the measured risk of the debtor. Thus the 8\% percentage 
was now calculated over the weighted amount and not over the total amount.  This version 
of the accord become effective at the beginning of the 2008 financial turmoil; with regulators under 
severe criticism from governments and the media, in 2010 the Committee issued a 
new version\cite{basel3} tightening capital rules.        

At the same time, since the beginning of the 2000s the academic community has been very 
critical of the capital rules, particularly because the VaR paradigm, 
on which such rules are based, assumes that returns are normally distributed 
and ``does not measure the distribution or extent of risk in the tail, but only provides 
an estimate of a particular point in the distribution''\cite{baselac}. In fact, there is a huge 
amount of evidence\cite{Borland_Bouchaud2005,Mandelbrot,Sornette,Bouchaud} that the 
returns of economic processes are not normally distributed, having typically heavy tails. 
According to the Central Limit Theorem\cite{Embrechts}, if returns are heavy-tail distributed, then
the underlying random variables have infinite variance or a variance of the order of the system 
size \cite{Mantegna_Stanley1994}. In economic systems, random variables are related to measurements 
taken from economic agents. Thus, the infinite variance results from long-range correlations 
between the economic agents. We will argue that this single fact compromises the stability of 
the flow and brings into question the effectiveness of capital level rules. 

Physics, and in particular statistical physics, has long inspired the construction of models 
for explaining the evolution of economies and societies and for tackling major economic 
decisions in different contexts\cite{samanidou,peinke}. The study of critical phenomena and 
multi-scale systems in physics led to the development of tools that proved to be useful in non-physical 
contexts, particularly in financial systems. One reason for this is that fast macroeconomic 
indicators, such as principal indices in financial markets, exhibit dynamical scaling, which is 
typical of critical physical systems\cite{Mantegna_Stanley1999}.
 
In this paper we will address the problem of the financial stability using statistical physics models 
that explain the occurrence of large crises, in order to show that the resilience of the banking 
system is not necessarily improved by raising capital levels. Our findings have a concrete social 
importance, since capital is the most expensive money a bank can provide to its debtors. Capital 
belongs to the shareholders, who bear the risk of the business and keep the job positions. So it 
must be remunerated above the money from depositors who do not bear these risks. Consequently, 
more capital means more costs on the flow of the money and, in the end, more constraints to economic development.   

We start in section \ref{sec:model} by describing an agent-based 
model\cite{samanidou} which enables us to generate the critical behaviour observed in economic systems. 
In section \ref{sec:macroproperties} we describe the observables that account for the economic properties of 
the system, namely the so-called overall product and business level\cite{tina}. 
Furthermore, the agent-based model as well as the macroscopic observables, are discussed for the 
specific situation of a network of banks and their
deposits and loans. One important property in financial
banking systems is introduced, namely the minimum capital level, defined
here through the basic properties of agents and their connections.
In section \ref{sec:results} 
we focus on the financial stability of the banking system, 
showing  that raising capital levels 
promotes concentration of economic agents if the economic 
production remains constant and it destroys economic production if that concentration does not 
occur. Finally, we present specific situations where each agent seeks the stability of its economic 
production after a raise in capital levels, leading to a state of worse financial stability, 
i.e.~a state in which large crises are more likely to occur. 
In section \ref{sec:conclusions} we draw the conclusions.

\section{Minimal model for avalanches of financial defaults}
\label{sec:model}

The model introduced in this section is based on a fundamental
feature that human beings have developed in their individual behaviour,
through natural selection, in order to be able to fight environmental 
threats collectively.  It is called {\it specialisation}\cite{lipseybook}, and
describes the tendency individuals have, when living in communities, to concentrate 
on one, or at most a few, specific tasks. 
Each individual does not need to do everything to survive, just to 
concentrate on a few tasks that he/she can do better for all the other
individuals. Everything else he or she will get from other specialised
elements of the community. 
Thus, specialisation leads to optimisation, enabling the entire community 
to accomplish goals otherwise unattainable. 
However, it also implies that individuals now need 
to exchange what they do, so that all have everything they need for 
survival. The set of all task and product exchanges between individuals
is what we usually call {\it Economy}.
Consequently, when building an economic system, a reasonable approach
is to take agents which are {\it impelled} to exchange some product
through a network of trades between pairs of agents.
In this scope, let us assume that the economic 
environment is composed of elementary particles called ‘agents’ 
and all phenomena occurring in it result from the interaction of 
those particles. Let us also assume that agents are attracted to 
interact, exchanging an observed quantity that takes the form of 
money, labour or other effective means used in the exchange. This 
type of model where the decision concerning an exchange is made 
by the exchanging agents alone is called a ``free-market economy''.

We represent these interactions or trades between agents through economic 
connections, and call the exchanged quantity `economic energy'.
Though the ``energy'' used here is not the same as physical energy, 
we will use the term in the economical context only.
Notice, however, that human labour is assumed to be ``energy'' delivered by
one individual to those with whom he/she interacts, which reward the 
individual with an energy that he/she accumulates. 
The balance between the labour (``energy'') produced for the neighbours
and the reward received from them may be positive (agent profits) or 
negative (agent accumulates debt). 
For details see Ref.~\cite{cruz_lind1}.
This analogy underlies the model introduced in the following, where
we omit the quotation marks and consider entities more general than
individual, which we call agents.
Agent-based models of financial markets have been intensively 
studied, see e.g.~Refs.~\cite{queiros2007,samanidou} and 
references therein.

Economic connections between two agents are in general not symmetric and 
there is one simple economic reason for that: if a connection were completely 
symmetric there would be no reason for each of the two agents to 
establish an exchange. In several branches of Economics we have different 
examples of these asymmetric economic connections like production/consumption, 
credit/deposit, a labour relation, repo’s, swaps, etc. 
In the next section, we will focus on a specific connection, namely
in credit/deposit connections.

Since each connection is asymmetric we distinguish the two agents involved by assigning two different 
types of economic energy. Hence, let us consider two connected agents, $i$ and $j$, where $i$ 
delivers to $j$ an amount of energy $W_{ij}$ and receives an amount $E_{ij} \neq W_{ij}$ in 
return. We call these connections the outgoing connections of agent $i$. The connections where 
agent $i$ receives from $j$ an amount of energy $W_{ji}$ and delivers in return $E_{ji}$ we 
call incoming connections.  

The energy balance for agent $i$ in one single trade connection is, from a labour 
production point of view, $U_{ij}=W_{ij}-E_{ij}$. 
Having two different types of energy, we choose $W_{ij}$ as the reference to which the other 
type $E_{ij}=\alpha_{ij}W_{ij}$ is related through the coefficient $\alpha_{ij}$ 
(see Eq.~(\ref{eq7prime}) below). Without loss of generality, we consider that one connection 
corresponds to the delivery of one unit of energy, $ W_{ij}=1$, yielding:
\begin{equation}
U _{ij}=1-\alpha_{ij} .
\label{eq:ibalance}
\end{equation}

In order to implement the model, we need to define the form of the coefficient $\alpha _{ij}$ in 
Eq.~(\ref{eq:ibalance}), which is a connection property. The amount of energy $E_{ij}$ that one 
agent $i$ gets in return for a delivered amount $W_{ij}$ can be taken as a price which depends 
on the rules of supply and demand. An agent delivering energy to many neighbours tends to impose 
a higher price on them. Similarly, an agent receiving energy from many other neighbours
will induce a reduction in the price imposed by its creditors. These principles can be incorporated in a 
simple {\it Ansatz} as:
\begin{equation}
\alpha_{ij}=\frac{2}{1+\hbox{e}^{-(k_{out,i}-k_{in,j})}}
\label{eq7prime}
\end{equation}
where $k_{out,i}$ is the number of outgoing connections of agent $i$ and $k_{in,j}$ is the number 
of incoming connections of neighbour $j$. For $\alpha_{ij}>1$ the energy provided by agent $i$ to 
agent $j$ is `paid' by $j$ above the amount of energy agent $i$ delivers. Thus, agent $i$ profits 
from this connection and gains a certain amount of energy, $U_{i}>0$. For $\alpha_{ij}<1$ the opposite occurs. 
From Eq.~(\ref{eq7prime}) one easily sees that $\alpha_{ij}$ is a 
step-function with average value one  and very small derivatives in the 
asymptotic limits $k_{in,j}\gg k_{out,i}$ and $k_{in,j}\ll k_{out,i}$.
Furthermore, in this latter limit $k_{in,j} \ll k_{out,i}$, the value 
of $\alpha_{ij}$ could in principle be any finite value 
larger than one. However, to guarantee symmetry  between the situation 
when agent $i$ profits from agent $j$, and that when agent $j$ profits from agent $i$, 
we consider the range $\alpha_{ij}\in [0,2]$, 
yielding $\alpha_{ij}=2$ for $k_{in,j} \ll k_{out,i}$.

Such energy transactions have an Economics analogue according to basic principles\cite{lipseybook}: 
a large (small) $k_{in,j}$ indicates a large (small) supply for agent $j$ and a large (small) $k_{out,i}$ 
indicates a large (small) demand of agent $i$. Thus, the difference $k_{out,i}-k_{in,j}$ measures the 
balance between the demand of an agent $i$ and the supply of its neighbour $j$.
$\alpha_{ij}$ saturates for large positive and negative differences in order to guarantee 
the price to be finite. 

The definition of $\alpha_{ij}$ in Eq.~(\ref{eq7prime}) is not uniquely determined 
by these economic requisites.
Similar functions such as the arc-tangent or the hyperbolic tangent have been 
used in this context\cite{bouchaud_arttghyper}. 
The main findings of this manuscript are not sensitive to the 
choice of functional dependency of $\alpha_{ij}$ as long as it is a 
step-function of $k_{in,j}-k_{out,i}$.

In the model described above, we disregard the economic details of agents and connections, keeping 
the model as general as possible. Still, this generalisation is not different in its essence from the 
one accountants must use to provide a common report for all sorts of business, with the difference that 
they use monetary units and we use dimensionless energy units. 

Because each agent typically has more than one neighbour, the total energy balance 
for agent $i$ is given by
\begin{equation}
U _{i}=\sum_{\hbox{\small all neighbours}}{U_{ij}} =\sum _{j \in \nu_{out,i} } (1-\alpha _{ij}) +  \sum _{j \in \nu_{in,i} } (\alpha _{ji}-1)  
\label{eq:itotbalance}
\end{equation}
where $j$ runs over all neighbours  of agent $i$, and 
$\nu_{out,i}$ and $\nu_{in,i}$ are, respectively the outgoing and incoming vicinities of the
agent.

This total energy balance $U_{i}$ is related to the well-known financial 
principle of net present value (NPV)\cite{matematica_financeira}: 
When an agent holds a deposit he or she supposedly pays for it 
(by definition) and most (but not all) accounting standards \cite{ref:IAS} 
assume it as a negative entry on the accounting balance. 
Here, we model deposits as a set of incoming connections from the same agent 
in which all associated cash-flows were already discounted.
In this way, if we could think of a balance sheet totally built with NPV's we 
would be near $U_{i}$. 

As we noted previously, economic energy is related to physical energy in the sense
that the agents must absorb finite amounts of physical energy from the environment to deliver 
economic energy.  Consequently, the economic energy balance $U_{i}$ of agent $i$ must be finite. 
The finiteness of $U_i$ for each agent is controlled by a threshold value, below which the agent 
is no longer able to consume energy from its neighbours, i.e.~below which it loses all its incoming 
connections. Furthermore, since this threshold reflects the incoming connections, it should depend 
on how many incoming connections our agent has. With such assumptions, 
we introduce the quantity 
\begin{equation}
c_i \equiv \frac{U_i}{\sum _{j \in \nu_{in,i}} (\alpha _{ji}-1)}
\label{eq1}
\end{equation}
for ascertaining if the agent is below a given threshold $c_{th}$ or not. 
We call this quantity $c_i$ the `leverage' of agent $i$.
Unlike we did previously in Ref.~\cite{cruz_lind1}, 
here $U_i$ is divided by the total product of the incoming 
connections solely and not by the `turnover'.
This choice is made to be in line with the way banking 
regulators define leverage. Still, this alternate definition does not change 
the critical behaviour observed in our model and previously 
reported\cite{cruz_lind1}. For the case that the mean-field approximation 
$\alpha_{ij}\sim \langle \alpha \rangle$ holds, the leverage $c_i$
depends exclusively on the network topology, yielding
$c_i=\tfrac{k_{out,i}}{k_{in,i}} - 1$\cite{cruz_lind1}.

Leverage has a specific meaning in Economics, which related to the quantity $c_i$:  
it measures the ratio between own money and total 
assets\cite{matematica_financeira}.  Thus, each agent has a 
leverage $c_i$ which varies in time and there is a threshold $c_{th}$ below 
which the agent `defaults' or goes bankrupt, losing its incoming connections with its neighbours.  
Since the bankrupted agent is connected to other agents, the energy balances must be updated for
every affected agent $j$. 
Bankruptcy leads to the removal of all incoming connections of 
agent $i$, reducing the consumption of the bankrupted agent to a minimum,
i.e.~keeping one single consumption connection, $k_{in,i}=1$.
This situation implies that agent $i$ and its neighbours $j$ should be updated as 
follows: 
\begin{subequations}
\begin{eqnarray}
c_{i}  & \rightarrow & k_{out,i}-1 \label{eq:avalanche2} \\
k_{out,j} & \rightarrow & k_{out,j}-1 \label{eq:avalanche3} \\
c_j  & \rightarrow & c_{j} - \frac{1}{k_{in,j}} .\label{eq:avalanche4}
\end{eqnarray} 
\label{eq:avalanche}
\end{subequations}
We keep the agent with one consumption connection in the system also to
avoid the divergence of $c_i$ as defined in the context of financial 
regulation\cite{basel1,basel2,basel3}. Such a minimum consumption value 
has no other effect on the problem we will be dealing with in the next section.

The bankruptcy of $i$ leads to an update of the energy balance for neighbour $j$, which may 
then also go bankrupt, and so on, thereby triggering a chain of bankruptcies henceforth called an 
`avalanche'. See illustration in Fig.~\ref{fig:avalanche}.

The concepts of leverage and leverage threshold are used by R.C. Merton\cite{merton74} and 
O. Vasicek\cite{MertonVasicek} in their credit risk models, which are the theoretical foundation 
for the Basel Accords\cite{basel1,basel2,basel3}. Namely, Merton assumed this threshold for pricing 
corporate risky bonds using a limit on debt-equity ratio and Vasicek generalised it to a 
``debtor wealth threshold'' below which the debtor would default on a loan.

\section{Macroproperties: overall product and business level}
\label{sec:macroproperties}

Let us consider a system of $L$ interconnected agents which form the environment where each agent 
establishes its trades. We call henceforth this environment the operating neighbourhood. We can measure 
the total economic energy of the system by summing up all outgoing connections to get the overall product 
$U_T$, namely
\begin{equation}
U_T=\sum _{i=1}^{N}{\sum_{j\in {\cal V}_{out,i}}} {(1-\alpha_{ij})},
\label{overall}
\end{equation}
where ${\cal V}_{out,i}$ is the outgoing vicinity of agent $i$, with $k_{out,i}$ neighbours. The quantity $U_T$ 
varies in time and its evolution reflects the development or failure of the underlying economy. Instead of $U_{T}$, 
we consider the relative variation $\frac{dU_{T}}{U_{T}}$, also known as `return' in a financial context.   
We can also measure the average business level per agent, defined as the moving average in time of the overall product:
\begin{equation}
\Omega=\frac{1}{L}\frac{1}{T_S}\int_{t}^{t+T_S} U_T(x)dx
\label{eq8}
\end{equation}
where $T_S$ is a sufficiently large period for taking time averages. Similar quantities are used in Economics as 
indicators of individual average standard of living\cite{tina}. In the continuum limit, the time derivative of the business 
level $\Omega$ gives the overall product uniformly distributed over all agents.

\begin{figure}[tb]
\centerline{\includegraphics[width=0.90\linewidth]{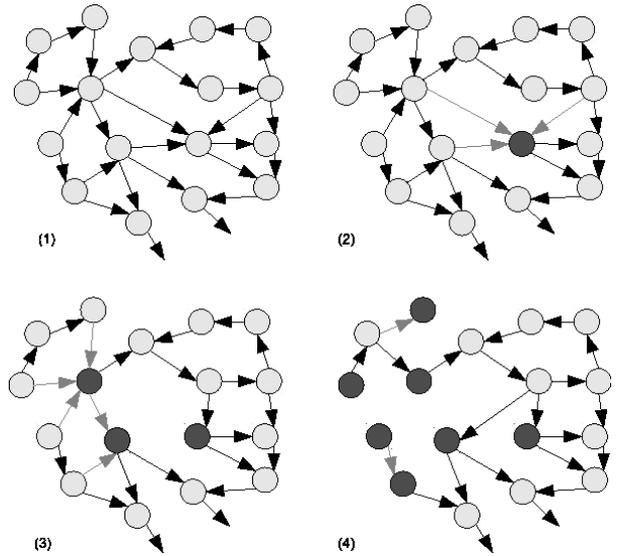}}
\caption{\protect
         Illustration of a bankruptcy avalanche. Each arrow points to the agent
	 for which it is an incoming connection. 
         Any agent in the economic environment
	 is part of a complex network {\bf (1)}, and 
         is susceptible to
          go bankrupt, 
	  which will destroy its incoming connections {\bf(2)}. Consequently, new energy balances 
	  must be updated for the affected neighbours, whose leverage can go over the threshold{\bf (3)}.
	  Since these neighbours also have neighbours of their own, connection will continue to be destroyed 
	  until all agents again have a leverage above the threshold. 
}
\label{fig:avalanche}
\end{figure}

At each time step a new connection is formed, according to the standard preferential attachment algorithm 
of Barab\'asi-Albert\cite{12}: For each connection created one agent is selected using a probability function 
based on its previous outgoing connections, expressed as 
\begin{equation}
P(i)=\frac{k_{out,i}}{\sum_{l=1}^{L}{ k_{out,l}}}
\label{eq:prefatt}
\end{equation}
and one other agent is selected by an analogous probability function built with incoming connections. Such a 
preferential attachment scheme is associated with power-law features observed in the Economy long ago\cite{pareto,hu06} 
and is here motivated by first principles in economics that agents are impelled to follow: an agent having 
a large number of outgoing connections is more likely to be selected again to have a new outgoing connection, 
and likewise for incoming connections.  

As connections are being created, a complex network of economic agents 
emerges and individual leverages 
(see Eq.~(\ref {eq1})) are changing until eventually 
one of the agents goes bankrupt ($c_i < c_{th}$) 
breaking its incoming connections and changing the leverage of its neighbours, who might also go bankrupt 
and break their incoming connections and so on. See Fig.~\ref{fig:avalanche}. This avalanche affects the total overall 
product, Eq.~(\ref{overall}), because the dissipated energy released during the avalanche is subtracted. This total 
dissipated energy is given by the total number of broken connections, and measures the `avalanche size', denoted
below by $s$. Since the avalanche can involve an arbitrary number of agents, and is bounded only by the size of the system, 
the distribution of the returns $\frac{dU_{T}}{U_T}$ will be heavy-tailed, as expected for an economic system. 
See Fig.~\ref{fig2} below.

Until now we have been dealing with generic economic agents that make generic economic connections between 
each other. No particular assumption has been made besides that they are attracted to each other to form 
connections by the mechanism of preferential attachment and that the economic network cannot have infinite 
energy. From this point onward, we will differentiate some of these agents, labeling them as `banks'. To this 
end we fix the nature of their incoming and outgoing connections: The incoming connections are called `deposits', 
the outgoing connections are called `loans'. We should emphasize that we are not singling out this kind of agent from 
the others. Banks are modelled as economic agents like any other. We have only named its incoming and outgoing connections, which
we could also do for all the remaining agents, as consuming/producing, salary/labour, pension/contribution, etc, to model 
every single business we could think of. We are choosing this particular kind of agent because banks are the object 
of banking regulation and the aim of financial stability laws.      

The threshold leverage $c_{th}$ for one bank represents its `minimum capital level'. The capital of one bank is 
really an amount of incoming connections, which are equivalent to deposits, because shareholders are also economic agents. 
This means that the `minimum capital level' in the model will be much higher than in real bank markets because we 
are disregarding shareholders and adding the remaining energy deficit to fulfill $c_{th}$. Therefore, we cannot 
map directly the levels obtained in the model onto the levels defined in banking regulation. We can, however, uncover 
the behaviour of economic agents in scenarios difficult to reproduce without such a model.

\section{Raising the minimum capital level}
\label{sec:results}

In this section we use the model described above in different scenarios, 
i.e.~for different sizes of the operating neighbourhood and different minimum 
capital levels. From Eq.~(\ref{eq1}) one sees that the leverage of one 
agent is always larger than $-1$. 
Since we deal with bankruptcy we are interested in negative values of 
$c_{th}$, which reduces the range of leverage values to [-1,0]. 
Our simulations showed that a representative range of values for both the 
threshold and the size $L$ of the operating neighbourhood is $[-0.72,-0.67]$ 
and $[500,2000]$ respectively. For each pair of values 
$\left(L, c_{th}\right)$ the system evolves until a total of 
$1.5\times 10^6$ connections are generated. 
We discard the first $10^5$ time-steps which are taken as transient. 
\begin{figure}[tb]
\centerline{\includegraphics[width=0.90\linewidth]{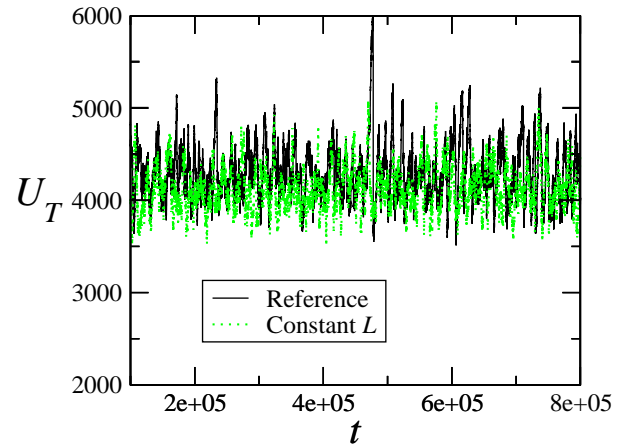}}
\caption{\protect
         (Colour online) 
         Illustration of the effect of raising the minimum capital level 
         on the overall product $U_T$, at constant $L=1500$. 
         Raising $c_{th}$ from $-0.71$ (solid line) to $-0.69$ 
         (dotted line) does not significantly change
         the overall product.
           }
\label{fig1_a}
\end{figure}
\begin{figure}[tb]
\centerline{\includegraphics[width=0.90\linewidth]{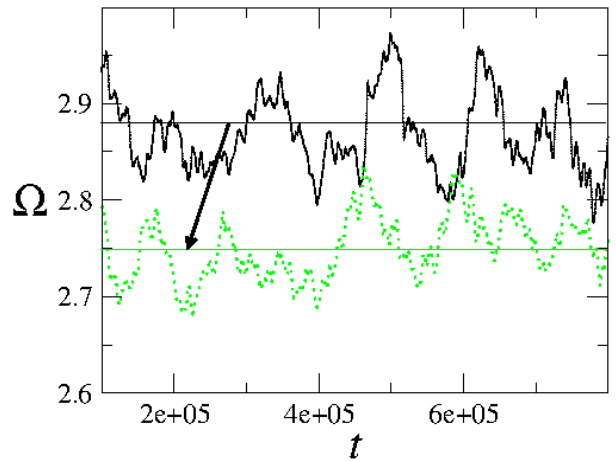}}
\caption{\protect
	  (Colour online)
          Illustration of the effect of raising the minimum capital on the business level at constant
	  $L=1500$. Raising $c_{th}$ from $-0.71$ (solid line) to $-0.69$ (dotted line) 
	  decreases the business level from $\Omega=2.88$ to $\Omega=2.75$. 
  }
\label{fig1_c}
\end{figure}

Figures \ref{fig1_a} and \ref{fig1_c} illustrate the evolution of the 
overall product $U_{T}$ and business level $\Omega$ for a situation in which
the minimum capital level is raised, while keeping the size of the operating neighbourhood constant. 
The solid line shows the initial situation with lower minimum capital level 
and the dashed line the final situation with higher minimum capital level. 
From Fig.~\ref{fig1_a} we can see that if the size of the 
operating neighbourhood is kept constant,
the quasi-stationary level of the overall product 
does not significantly change. 

Following this observation we next
investigate the evolution of the 
return distribution for $U_T$,
considering an increase of the minimum capital 
level at constant size $L$ of the operating neighbourhood. 
To this end we compute the cumulative size distribution of 
avalanches, i.e.~the fraction $P_{c}(s)$ of avalanches of size larger 
than $s$.
Numerically, the size $s$ of an avalanche is found by summing all connections 
destroyed during that avalanche. The value of $P_c(s)$ is then obtained by identifying  
the avalanches whose size is greater than $s$.
\begin{figure}[tb]
\centerline{\includegraphics[width=0.90\linewidth]{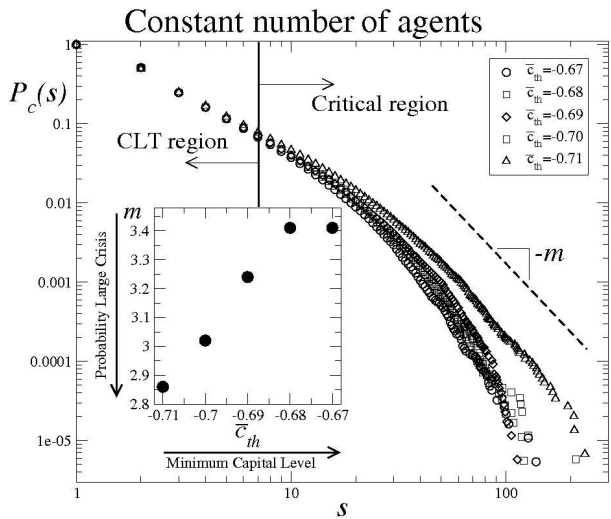}}
 \caption{\protect 
        Avalanche (crises) size distributions for different scenarios of 
        minimum capital level, keeping the operating neighbourhood unchanged for 
        each agent. 
        The different distributions match at small sizes, in the 
        region where the Central Limit Theorem (CLT) holds, 
        and deviate from each other for larger crises (critical region). 
        In the critical region one observes (inset) that increasing the 
        minimum capital level decreases the probability for a large avalanche
        to occur, which supports the intentions of the Basel III accords. However, 
        in this scenario one assumes that each bank will have a simultaneous 
        decrease of their business level (see text and Fig.~\ref{fig1_c}). 
        A more natural scenario would be one where each bank reacts to 
        the rise in the minimum capital level in such a way as to keep 
        its business level constant, which leads to a completely different 
        crises situation (see Fig.~\ref{fig3}).}
\label{fig2}
\end{figure}

Figure \ref{fig2} shows the cumulative size distribution of avalanches for 
different minimum capital levels, keeping $L=2000$. For small avalanche 
sizes, the Central Limit Theorem holds\cite{Mantegna_Stanley1994} and thus all 
size distributions match independently of the minimum capital level. 
For large enough avalanches (`critical region'), 
the size distributions deviate from each other, exhibiting a power-law tail 
$P_c(s)\sim s^{-m}$ with an exponent $m$ that depends on the minimum capital 
level $c_{th}$ (inset). 
As expected\cite{cruz_lind3},
the exponent found for the avalanche size distribution takes 
values in the interval $2 < m < 7/2$.

As can be seen in the inset of Fig.~\ref{fig2} the exponent increases in 
absolute value for larger minimum capital levels, indicating a smaller
probability for large avalanches to occur. 
However, this scenario occurs only when the size 
of the operating neighbourhood is kept constant
and, as shown in Fig.~\ref{fig1_c}, 
the increase of the minimum capital level is also accompanied by a decrease of 
the business level. This means that each agent has 
less economic energy or, in current language, is poorer. 

\begin{figure}[tb]
\centerline{\includegraphics[width=0.90\linewidth]{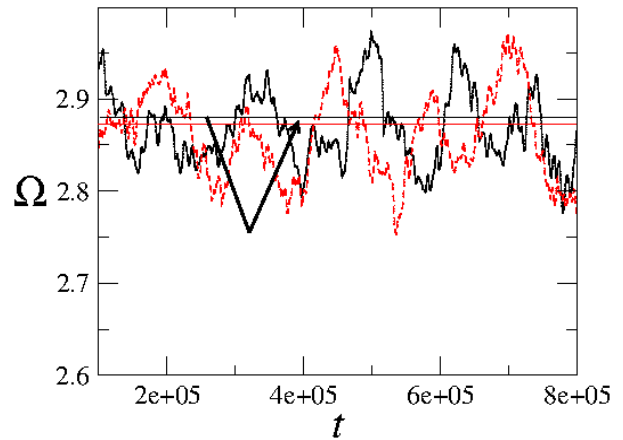}}
\caption{\protect
	  (Colour online)
         Unlike in Figs.~\ref{fig1_a} and \ref{fig1_c} it is 
         possible to raise the minimum capital level $c_{th}$ 
	 from $-0.71$ (solid line) to $-0.69$ (dashed line), while keeping 
         the business level constant.
         In the case plotted, $\Omega\sim 2.88$}.
\label{fig1_d}
\end{figure}

\begin{figure}[t]
\centerline{\includegraphics[width=0.90\linewidth]{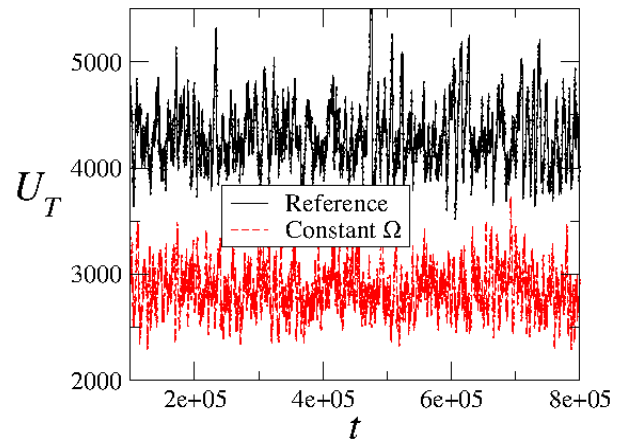}}
\caption{\protect
	(Colour online)
        Keeping the business level constant at $\Omega\sim 2.88$ and 
        raising the minimum capital level from $-0.71$ 
	(solid line) to $-0.69$ (dashed line) leads to a decrease of the 
        operating neighbourhood, which is reflected in a lower overall product.}
\label{fig1_b}
\end{figure}

Assuming that agents do not want to be poorer despite regulatory constraints, 
and therefore try  
to keep their business levels constant (Fig. \ref{fig1_d}), a natural 
reaction against raising the minimum capital level is to decrease 
the number of neighbours with whom the agent establishes trade connections, 
i.e. to decrease the size of the operating neighbourhood (Fig. \ref{fig1_b}). 
In Economics this is called a concentration process\cite{ref:sox}, which 
typically occurs when the regulation rules are tightened up. In such a scenario 
where the size of the operating neighbourhood is adapted so as to 
maintain the business level constant, the distributions plotted 
in Fig.~\ref{fig2} are no longer observed. In particular, the exponent $m$ 
does not increase monotonically with the minimum capital level
as we show next. 
\begin{figure}
\includegraphics[width=8.5cm,angle=0]{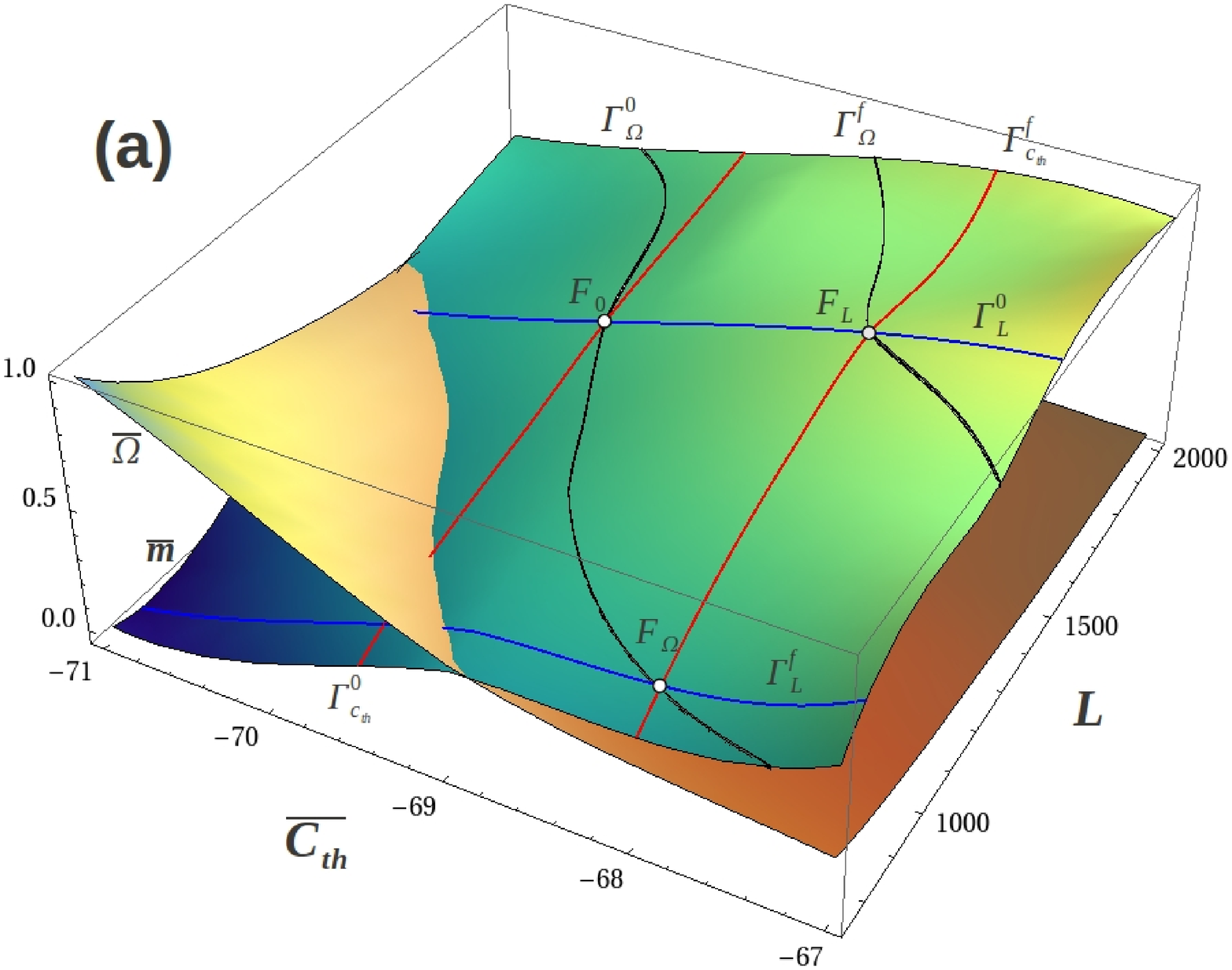}
\includegraphics[width=8.5cm,angle=0]{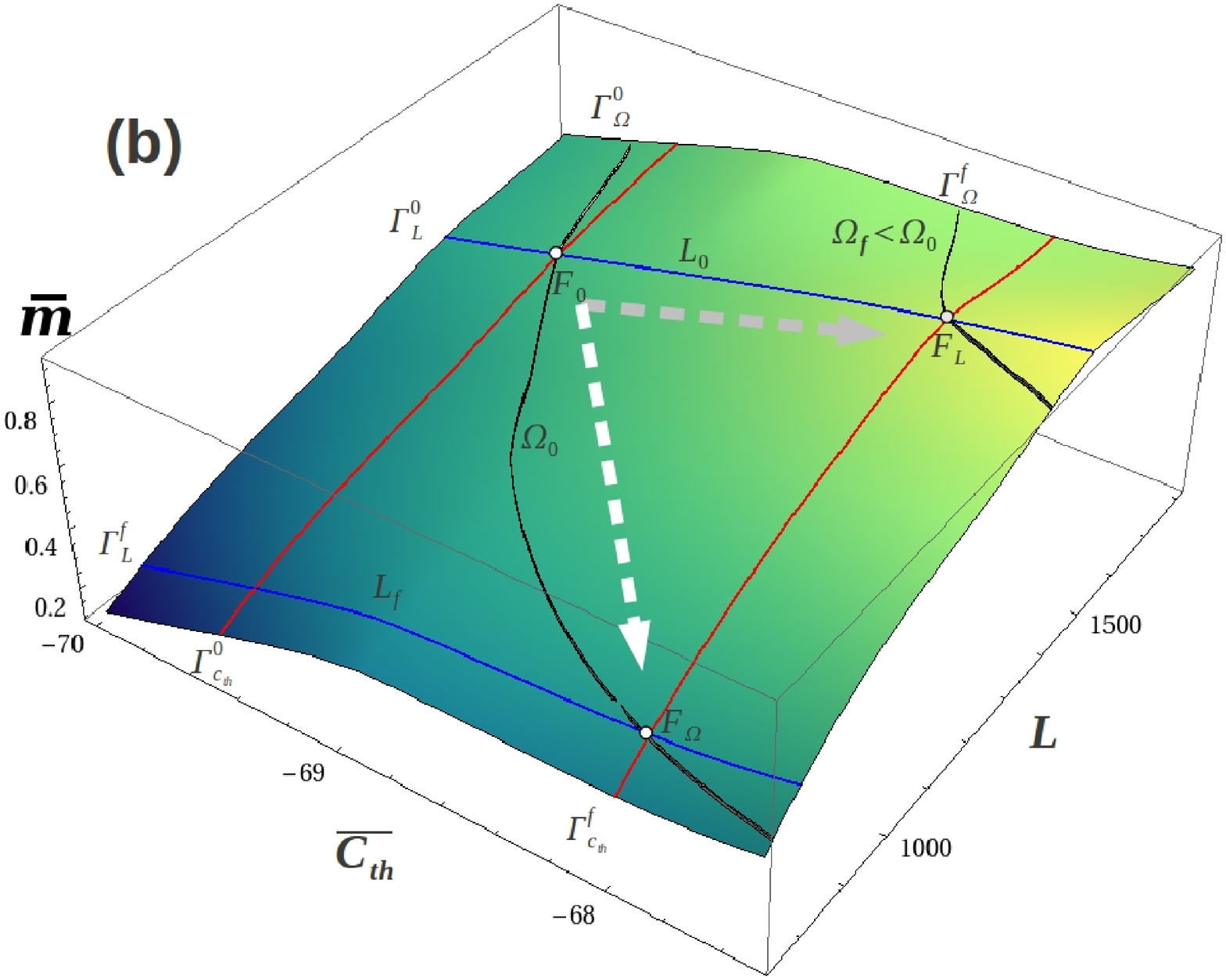}
\caption{\protect 
	  (Colour online)
         {\bf (a)} Normalized critical exponent $\bar{m}$ and business 
         level $\Omega$ as functions of the minimum capital level 
         $c_{th}$ and system size $L$.  
         For an initial financial state $F_0$, an increase of the minimum 
         capital level takes the system along one of the infinitely many 
         paths between the initial and final isolines at constant minimum 
         capital level, $\Gamma_{c_{th}}^0$ and $\Gamma_{c_{th}}^f$ respectively.
         {\bf (b)} If such a path follows the isoline at constant system 
         size, $\Gamma_L^0$, the critical exponent increases and thus the
         probability for large avalanches decreases. Simultaneously however, 
         its business level decreases 
         ($\Omega_f < \Omega_0$), which runs against the natural intentions 
         of financial agents. 
         On the contrary, if the path is along the isoline  at constant 
         business level, $\Gamma_{\Omega}^0$, as one naturally 
         expects the financial agents would do, the critical exponent does 
         not change significantly, meaning that large financial crises may 
         still occur with the same probability as before (see text).}
\label{fig3}
\end{figure}

Figure \ref{fig3}a shows the critical exponent $m$ and the business level 
per agent $\Omega$ as functions of the minimum capital level $c_{th}$ 
and the operating neighbourhood size $L$. For easy comparison, both quantities 
are normalized in the unit interval of accessible values. 

The critical exponent shows a tendency to increase with both the 
minimum capital level and the operating neighbourhood 
size. The business level, on the other hand, decreases when the 
minimum capital level or the neighbourhood size increase.
Considering a reference state $F_0$ with $c_{th,0}$, $L_0$ and $\Omega_0$ 
there is one isoline of constant minimum capital level, $\Gamma_{c_{th}}^0$, 
and another of constant operating neighbourhood size, $\Gamma_{L}^0$, crossing 
at $F_0$. Assuming a transition of our system to a larger minimum capital 
level at isoline $\Gamma_{c_{th}}^f$ while keeping $L$ constant, i.e. along the 
isoline $\Gamma_L^0$, one arrives at a new state $F_L$ with a larger critical 
exponent, which means a lower probability for large avalanches to occur, as 
explained above. However in such a situation the new business level $\Omega_f$ 
is lower than the previous one $\Omega_0$. 

On the contrary, if we assume that the transition from $F_0$
to the higher minimum capital level occurs at constant business level, 
i.e. along the isoline $\Gamma_{\Omega}^0$, one arrives to a state $F_{\Omega}$ 
on the isoline $\Gamma_{c_{th}}^f$ for which the critical exponent is 
not necessarily smaller than for the initial state.

From economical and financial reasoning, one typically assumes that, 
independently of external directives, under unfavourable circumstances
economical and financial agents try, at least, to maintain their business 
level. This behaviour on the part of agents leads to a situation which contradicts the 
expectations of the Basel accords and raises the question of whether such regulation 
will indeed prevent larger avalanches from occurring again in the future. 
To illustrate this, Fig.~\ref{fig3}b shows a close-up of the 
$m$-surface plotted in Fig.~\ref{fig3}a. 

For the reference state $F_0$ one finds an exponent $m=2.97\pm 0.18$. 
An increase of the minimum capital level at 
constant operating neighbourhood size (state $F_L$) yields $m=3.34\pm 0.09$,
while increasing the minimum capital level at constant business 
level (state $F_{\Omega}$), yields $m=2.79\pm 0.09$, which corresponds to a 
significantly higher probability that large avalanches will occur.

\section{Discussion and conclusion}
\label{sec:conclusions}

In summary, raising the minimum capital levels may not necessarily improve 
banking system resilience. Resilience may remain the same if banks go after the 
same business levels, as one should expect, according to economic reasoning. 
Indeed, since business levels are part of the achievement of any economic 
agent that enters a network of trades, each agent will try, 
at least, to maintain this level, independently of regulatory constraints.
 
Furthermore, our findings can solve the apparent contradiction between the credit risk 
models that serve as the theoretical foundation for bank stability accords 
and the definition of capital levels. In fact, bank stability accords 
impose on banks an adapted version of Merton-Vasicek 
model\cite{MertonVasicek} in which it is assumed that each agent has a 
leverage threshold above which it defaults on credit. The assumption of 
this threshold combined with a first principle of Economics 
-- that the Economy emerges from the exchanges between agents -- naturally leads  
to an interplay between agents that can propagate the effect of one 
default throughout the entire economic system.

Economic systems have long-range correlations and heavy-tailed 
distributions that are not compatible with a linear assumption that 
raising individual capital levels will lead to stronger stability. 
Because of the interdependency, this assumption is probably valid only
in two situations: when it is impossible for an individual to default; and 
when individuals behave independently from each other (random trade
connections). Both situations do not occur in real economic systems.

These findings can inform the recent governmental measures for 
dealing with the effects of the 2008 financial crises. 
In particular, governments have shown\cite{basel3} a tendency for imposing 
a higher capital investment from banks.  If the 
threshold is increased, while the total amount of trade remains constant, there 
will be fewer trade connections between the 
banks and their clients, which leads to smaller avalanches in the evolution of the 
financial network. 
On the other hand, if the total amount of trade is assumed to grow, following
the rise in minimum capital, the probability of greater 
avalanches will also increase to the level where it was before or even to a higher level.


The scale-free topology of the economic network plays a major role in the 
determination of the size distribution of the avalanches. 
At the same time, the scale-free topology emerges naturally from the rules
introduced, which are motivated by economic reasoning, namely the principles of 
demand and supply. 
Still, one could argue that for bank regulation purposes, a different
(imposed) topology for the connections between financial agents would 
help to prevent large crises. For example, if the economic 
network is structured as a random Erd\"os-R\'enyi network\cite{erdos}, 
in which every economic agent has 
the same probability of being chosen to form an economic connection, the system 
would not have avalanches. In such a model, since connections are equally 
distributed throughout the system, all agents would have statistically the 
same balance. 
In other words, for each bankruptcy the expected number of child bankruptcies 
in the avalanche would have either zero size or the size of the system.
Thus, with Erd\"os-R\'enyi topology, one expects still the danger of 
triggering such a large chain of insolvencies able to collapse the entire 
system.

Directives more oriented to the connection topology emerging in the financial network could 
be a good alternative. Interestingly, although controversial, our claims point in the direction 
of IMF reports in November 2010\cite{11}, where it is argued 
that rapid growth in emerging economic periods can be followed by financial crises, and also to 
recent theoretical studies on the risk of interbank markets\cite{13,14}. Indeed the recent IMF Memorandum 
on Portuguese economic policy\cite{imf} already includes directives that 
reveal IMF's concern not only with tuning capital buffers and other local  
properties but also with monitoring the banking system as a whole, and in particular keeping track of 
the financial situation of the largest banks in the network. We believe that such global networking
measures are much more trustworthy than local ones.

\section*{ACKNOWLEDGMENTS}
The authors thank M.~Haase, F.~Raischel, P.I.~Teixeira and J. Jarego for
useful discussions and for the reading of the ma\-nus\-cript.
The authors acknowledge financial support from PEst-OE/\-FIS/\-UI0618/\-2011, 
PGL thanks  {\it Funda\c{c}\~ao para a Ci\^encia e a Tecnologia -- 
Ci\^encia 2007} for financial support.


\end{document}